# Micromagnetic understanding of stochastic resonance driven by spin-transfer-torque

G. Finocchio[1,*], I. N. Krivorotov[2], X. Cheng,[2] L. Torres[3], B. Azzerboni[1]

[1]Dipartimento di Fisica della Materia e Ingegneria Elettronica, University of Messina, Salita Sperone 31, 98166 Messina, Italy.

[2]Department of Physics and Astronomy, University of California, Irvine, 92697-4575 CA, USA.

[3]Departamento de Fisica Aplicada, University of Salamanca, Plaza de la Merced s/n, 37008 Salamanca, Spain.

## ABSTRACT

In this paper, we employ micromagnetic simulations to study non-adiabatic stochastic resonance (NASR) excited by spin-transfer torque in a super-paramagnetic free layer nanomagnet of a nanoscale spin valve. We find that NASR dynamics involves thermally activated transitions among two static states and a single dynamic state of the nanomagnet and can be well understood in the framework of Markov chain rate theory. Our simulations show that a direct voltage generated by the spin valve at the NASR frequency is at least one order of magnitude greater than the dc voltage generated off the NASR frequency. Our computations also reproduce the main experimentally observed features of NASR such as the resonance frequency, the temperature dependence and the current bias dependence of the resonance amplitude. We propose a simple design of a microwave signal detector based on NASR driven by spin transfer torque.

# I. INTRODUCTION

Spin transfer torque (STT) [1,2,3] can give rise to magnetization self-oscillations,[4,5,6,7,8,9,10] and magnetization reversal[11,12,13] in magnetic nanocontacts, nanoscale spin valves and magnetic tunnel junctions. The main potential applications of STT are magnetic random access memories, microwave nano-oscillators, frequency modulators and microwave signal detectors.[14,15] Alternating STT can induce non-linear phenomena such as injection locking,[16] non-linear frequency modulation,[17] and resonant switching of magnetization.[18] Ac STT applied to a spin valve can also generate a dc voltage (diode effect) due to ferromagnetic resonance (FMR) driven by STT (ST-FMR).[14] A recent experiment have shown that for super-paramagnetic nanomagnets, the combined action of ac-STT and thermal fluctuations can give rise to adiabatic and non-adiabatic stochastic resonance of magnetization that significantly enhance the dc voltage generated in response to ac STT.[19,20] In this paper, we make finite-temperature micromagnetic simulations in order to elucidate the origin of the nonadiabatic stochastic resonance (NASR).

Stochastic resonance has been observed in a broad range of systems.[21,22,23,24] It occurs in the presence of a Gaussian noise, a "weak" periodic force, and an energy landscape with a few energy minima, in which the energy separation among the minima and the noise temperature are of the same order of magnitude.[21] For over-damped systems, a low-frequency weak periodic drive transforms random thermally-activated transitions among the energy minima into quasi-periodic transitions with the frequency of the weak ac drive. Here, we performed a systematic micromagnetic study to elucidate the mechanism of stochastic resonance of magnetization in exchange biased nanoscale spin-valves driven by weak alternating STT. In our numerical experiments, we find a regime of a low-frequency (adiabatic) stochastic resonance (ASR), for which we observe quasi-periodic hopping of the free layer magnetic moment between a static state (anti-parallel state of the spin valve) and a dynamic state of self-oscillations. The hopping is synchronized with the frequency of a low-frequency (50 MHz) alternating current that applies ac STT to the free layer. We also find a regime of NASR in which transitions among two static (parallel and anti-parallel states of the spin valve) and a dynamical state are observed under the action of a high-frequency (GHz-range) ac STT drive. Our results suggest that thermally-activated transition in the NASR regime are described by a regular Markov chain with different transition paths.

Our computations reproduce the main features observed in the experiment[19] such as the resonance frequency, the temperature dependence of the resonance amplitude at a fixed current and the current dependence of the resonance amplitude at a fixed temperature. In particular, at the NASR frequency, we observe enhancement of the dc voltage generated by the spin valve in response to ac STT by two orders of magnitude compared to the dc voltage observed in standard ST-FMR measurements. We argue that NASR in spin valves can find applications in sensitive microwave signal detection and can be used for signal demodulation in the frequency-shift keying (FSK) demodulation scheme.[25,26]

The paper is organized as follows. Section II introduces the micromagnetic framework of our numerical experiment. Section III describes results of the micromagnetic behaviour of the system studied: dynamics induced by dc STT, the effect of thermal fluctuations on the dynamics, and simulations of zero-temperature ST-FMR. Finally, micromagnetic simulations of SR are described in Section IV.

## II. MICROMAGNETIC FRAMEWORK DESCRIPTION

We study, by means of full micromagnetic simulations, the magnetization dynamics of a nanoscale spin valve with the same material and geometry parameters as the system experimentally studied in Refs. [19, 20]. The active part of the spin-valves (see the inset of Fig.1(a)) consists of a Py(3 nm)/ Cu(6 nm)/ Co(3 nm)/ $Ir_{20}Mn_{80}$(8 nm) (Py=$Ni_{81}Fe_{19}$) multilayer, where the Py and the Co layers act as the "free" and "reference" layers respectively. The spin valve is patterned into an elliptical 120×60 $nm^2$ nanopillar. We use a Cartesian coordinate system where the $x$ and the $y$ axes are the long and the short in-plane axes of the ellipse, we refer to **m** and **$m_p$** as normalized magnetization of the "free" and "reference" layers respectively. A 50 mT exchange bias field applied to the Co nanomagnet from $Ir_{20}Mn_{80}$ is directed along the long (easy) axis of the ellipse (positive x-axis). We use the system parameters consistent with the experiment[19]: for the Py-layer we use saturation magnetization $M_s$ = 650×$10^3$ A/m, exchange constant 1.3×$10^{-11}$ J/m, Gilbert damping $\alpha$ = 0.025; for the Co-layer we use saturation magnetization of 2000×$10^3$ A/m, exchange constant of 1.4×$10^{-11}$ J/m and Gilbert damping of 0.20. The Gilbert damping of the Co layer is large due to damping enhancement by exchange bias[27]. We characterize the quasi-static behaviour of the device magnetoresistance versus external magnetic field by means of self-consistent 3D micromagnetic simulations.[28] Fig. 1(a) shows the hysteresis loop of the average x-component of the Py layer magnetization as a function of field for two field directions: (i) positive x-direction (dashed line) and (ii) positive z-direction (out-of-plane) with a tilt angle of 10° along the x axis (dotted line). The latter field direction is used for the observation of stochastic resonance. For this latter field direction, we also performed simulations of magnetization dynamics driven by STT. In the rest of the paper, we present results for a field (*H*) of 200 mT; qualitatively similar results were obtained for *H* between 190 mT and 220 mT.

Our simulations of magnetization dynamics are based on the numerical solution of the Landau-Lifshitz-Gilbert-Slonczweski[1]:

$$\frac{d\mathbf{m}}{d\tau} = -(\mathbf{m} \times \mathbf{h}_{eff}) + \alpha \ \mathbf{m} \times \frac{d\mathbf{m}}{d\tau} - \frac{g}{e\gamma_0} \frac{|\mu_B| J}{M_s^2 d} \varepsilon(\mathbf{m},\mathbf{m_p}) \ \mathbf{m} \times (\mathbf{m} \times \mathbf{m_p}) \qquad (1)$$

where $g$ is the gyromagnetic splitting factor, $\gamma_0$ is the gyromagnetic ratio, $\mu_B$ is the Bohr magneton, $J$ is the current density, $d$ is the thickness of the free layer, $e$ is the electron charge, $d\tau = \gamma_0 M_S dt$ is the dimensionless time step, and $\varepsilon(\mathbf{m},\mathbf{m_p})$ characterizes the angular dependence of the Slonczewski spin torque term. By convention, positive current polarity corresponds to electron flow from the free to the pinned layer of the spin valve. $\mathbf{h}_{eff}$ is the dimensionless effective field. For the STT, we use the form derived by Slonczweski[29] with polarization $P=0.38$ and STT angular asymmetry parameter $\chi=1.5$, $\varepsilon(\mathbf{m},\mathbf{m_p}) = 0.5P(\chi+1)/(2+\chi-\chi\cos(\theta))$, where $\theta$ is the angle between $\mathbf{m}$ and $\mathbf{m_p}$. For the dynamical simulations, we assume $\mathbf{m_p}$ to be immobile after the computation of its static spatial configuration via a 3D micromagnetic simulations.[30] This is a reasonable approximation because the Co layer is much less sensitive to spin torque due to a larger value of the damping parameter[27] and due to larger saturation magnetization. In the simulations, the effective field $\mathbf{h}_{eff}$ acting on the Py magnetization includes all standard micromagnetic contributions (external, exchange, magnetostatic), the magnetostatic coupling to the Co-layer and the Oersted field from the current. A complete description of the micromagnetic dynamical model is given in Refs. [30,31].

The noise induced by the thermal fluctuations in the magnetic system is modelled as an additive stochastic field $\mathbf{h}_{th}$ added to the deterministic effective field in each computational cell. The $\mathbf{h}_{th}$ is a three-dimensional vector quantity given by $\mathbf{h}_{th} = \frac{\xi}{M_s}\sqrt{D} = \frac{\xi}{M_s}\sqrt{\frac{2\alpha k_B T}{\mu_0 \gamma_0 \Delta V M_s \Delta t}}$ where $k_B$ is the Boltzmann constant, $\Delta V$ and $\Delta t$ are the discretization volume and the integration time step respectively, $T$ is the sample temperature, $\xi$ is a three-dimensional white Gaussian noise with zero mean and unit variance, it is uncorrelated for each computational cell.[32,33]

The current density is assumed to be uniform in the cross-sectional area of the spin-valve. The magneto-resistance is computed as an average over the cross-sectional area $r(\mathbf{m_p}, \mathbf{m_f}) = \frac{1}{N_f} \sum_{i=1...N_f} r_i(\mathbf{m}_{i,\mathbf{p}}, \mathbf{m}_{i,\mathbf{f}})$, where $N_f$ is the number of computational cells and $r_i(\mathbf{m}_{i,\mathbf{p}}, \mathbf{m}_{i,\mathbf{f}})$ is the magneto-resistance of the $i^{\text{th}}$ 1 cell computed as $r_i(\mathbf{m}_{i,\mathbf{p}}, \mathbf{m}_{i,\mathbf{f}}) = 0.5[1 - \cos(\theta_i/2)]/[1 + \chi \cos(\theta_i/2)]$ ( $\cos(\theta_i/2) = 0.5(1 + \mathbf{m}_{i,\mathbf{p}} \bullet \mathbf{m}_{i,\mathbf{f}})$ ).

## III. NUMERICAL CHARACTERIZATION OF THE SYSTEM

As discussed in the introduction, the necessary ingredients for the observation of stochastic resonance are coexistence of at least two energy minima in the energy landscape, a weak periodic drive and a source of noise.[21] For the magnetic field direction studied here (200 mT) and without thermal fluctuations, we determine the interval of direct current densities, in which two states of the Py layer magnetization coexist: (i) a dynamic state of magnetization self-oscillations, D, and the static state corresponding to nearly anti-parallel (AP) configuration of the spin valve. Figure 1(b) summarizes our results by showing the hysteresis loop (oscillation frequency versus the direct current density $J$) obtained by sweeping the current density from $J_{dc} = 0$ to $J_{dc} = 0.7 \times 10^8$ A/cm$^2$ and back. At zero current density, the initial configuration of the magnetization is the AP state. For current densities $J_{dc} < J_{ON}$, the free layer magnetization is found to be in a stable static state (near the AP configuration) by numerically solving a generalization of the static LLGS equation (a generalization of the Brown equation, Eq. 1 with $\frac{d\mathbf{m}}{d\tau} = 0$ and $J_{dc} \neq 0$). In this static state, the free layer magnetization **m** and the generalized effective field are parallel for each computational cell $\mathbf{m} // (\mathbf{h}_{eff} - \sigma J_{dc} \varepsilon(\mathbf{m},\mathbf{m_p})\,(\mathbf{m} \times \mathbf{m_p}))$, where

$$\sigma = \frac{g\,|\mu_B|}{|e|\gamma_0\,M_s^2\,d}.$$

The state of magnetization auto-oscillations (the D state) is excited at a critical value of the current density $J_{ON} = 0.63\ 10^8$ A/cm$^2$. The D-state is characterized by a single-mode with an oscillation frequency of approximately 4 GHz and a significant oscillation power. The micromagnetic spectral mapping technique[34] shows that the excited mode is spatially uniform. Once the D state is excited, it is stable for sub-critical values $J_{dc} < J_{ON}$ down to a $J_{OFF} = 0.32 \times 10^8$ A/cm$^2$ (oscillation frequency around 2.4 GHz). For $J_{dc} < J_{OFF}$, the D state disappears and the only stable magnetization state is the static AP state. Fig.1 (c) shows that the D state is characterized by a blue shift of oscillation frequency as function of power of the oscillatory mode (computed as non-linear power[35]), which is typical of the out-of-plane mode of magnetization oscillations. The inset of Fig. 1(c) shows an example of the trajectory of the magnetization in the D state computed for $J_{dc} = 0.43 \times 10^8$ A/cm$^2$. This trajectory is nearly circular.

Even though STT is non-conservative, arguments based on the fluctuation-dissipation theorem[36] and experiments[37] can be invoked to describe the effect of STT as variation of the energy barrier between two states with well-defined energies, in our case the AP and the D state. With this approximation in mind, a bi-stable system can be used as an oversimplified model describing the magnetic behaviour of the spin-valve in the range of current densities where the D and the AP states coexist (see the insets in Fig. 1(b)). In this model, an energy barrier separates the D and the AP states, which correspond to two energy minima of the potential. The potential is asymmetric with the AP and D energy minima and the barrier separating them dependent on the current density $J_{dc}$ [37]. In particular, the D state is energetically favourable near $J_{ON}$, while the AP state is energetically favourable near $J_{OFF}$. These considerations will be important in the discussion of the simulation results in the presence of thermal fluctuations. Analytical calculations show the origin of the D/AP bi-stability region results from the coexistence of a fixed point of dynamics and a limit cycle in the dynamical stability diagram.[38]

In the study of SR driven by ac STT, we use as a weak periodic STT drive applied through an ac microwave current density $J_{AC} = J_M \sin(2\pi f_{AC} t)$ ($f_{AC}$ is the alternating current frequency), where $J_M \leq 0.1(J_{ON} + J_{OFF}) \leq 0.1 \times 10^8 \text{A/cm}^2$. To compare the simulation results to the experiment[19], we study the behaviour the system in terms of dc voltage generated by the spin valve in response to the applied ac current under direct current density bias in the range $J_{OFF} < J_{dc} < J_{ON}$. In order to directly compare our results with the experiment[19], we calculate the induced direct voltage ($V_{dc}$) as mean value of the product between the magneto-resistance and the current:

$$V_{dc} = mean((J_{dc} + J_{AC})S\Delta R_0 r(\mathbf{m_p},\mathbf{m_f})) - V_{dc-g} \qquad (2)$$

where S is the cross sectional area of the nanopillar, $V_{dc-g}$ is the dc background voltage due to the dc current, and $\Delta R_0$ is the difference between the resistance in the anti-parallel and the parallel configurations of the spin valve ($\Delta R_0 = 40 m\Omega$ for our system). $V_{dc-g}$ is subtracted for a direct comparison to the experiment, in which the dc background is automatically subtracted by the lock-in detection technique.[19,20]

Fig. 1(d) shows the calculated FMR frequency of the free layer as a function of the direct current density in the AP state computed for $J_M = 0.03 \times 10^8$ A/cm² (see Ref. [39] for a complete description of the numerical method). The inset of Fig. 1(d) shows an example of the calculated ST-FMR spectrum ($J_{dc}= 0.43 \times 10^8$ A/cm²). Using Eq. (2) we estimated, for the ST-FMR spectra, a maximum dc voltage obtained for a $J_M = 0.1 \times 10^8 \text{A/cm}^2$ around $0.3 \mu V$, this value is consistent with the results obtained using the simplified formula $V_{dc} = 0.5 J_M S \Delta R_{MAX} \cos(\beta_{FMR})$, $\Delta R_{MAX}$ and $\beta_{FMR}$ being the amplitude of the oscillating magnetoresistance at the FMR frequency and the phase between the oscillating magneto-resistance and the $J_{AC}$ respectively.

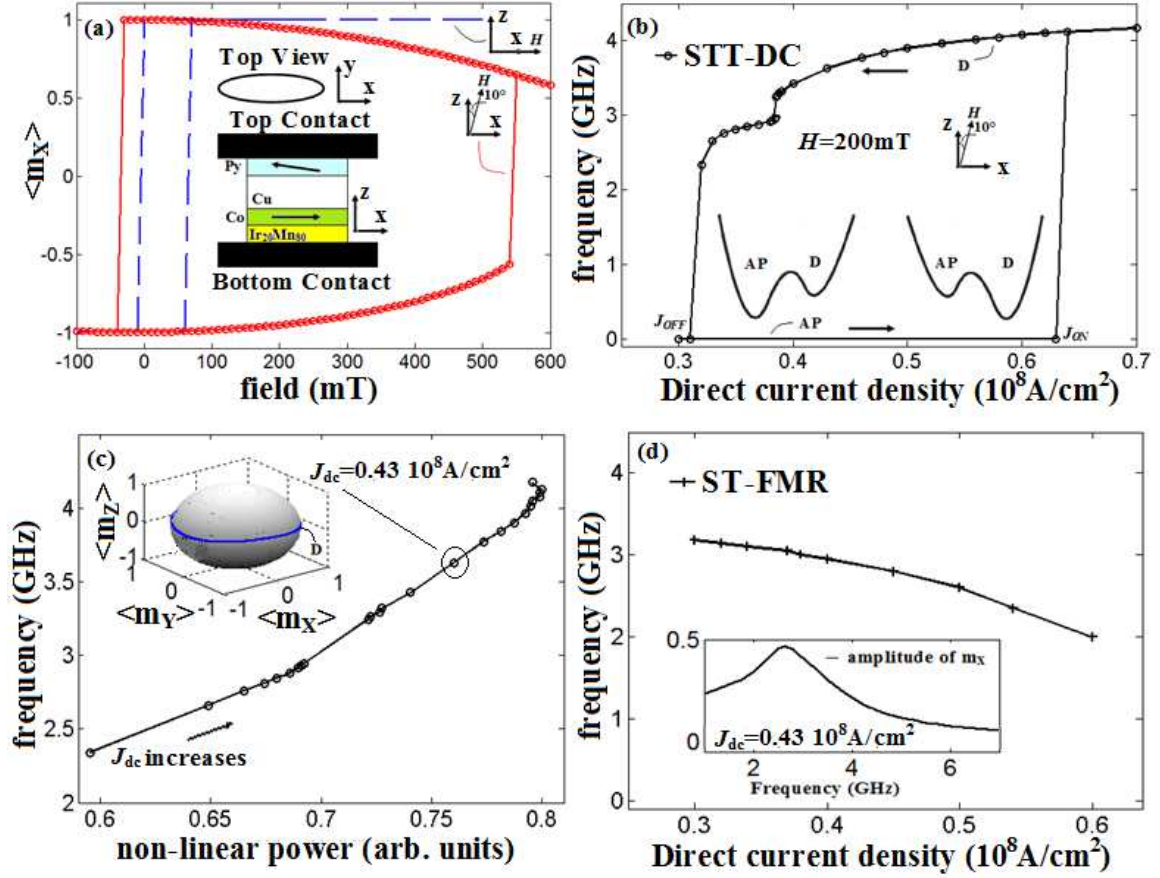

FIG. 1: (Color online) (a) Resistance-field hysteresis loops of the average x-component of the Py layer magnetization for two magnetic field directions: (i) field applied along the x-direction (dashed line) (ii) field applied along the z-direction with a tilt angle of 10° towards the positive x-direction (dotted line). The inset shows a sketch of the device and the Cartesian coordinate system. (b) Frequency of auto-oscillations of the Py layer at $H = 200$ mT in the configuration (ii) defined in Fig. 1 (a), for a range of current densities where the D and AP states coexist. The insets show magnetic energy landscape for a simplified model of the system described in the text. (c) Frequency-power relation for the D-state. Inset: magnetization trajectory in the D-state on the unit sphere. (d) Calculated ST-FMR frequency in the AP state versus direct current density $J_{OFF} < J_{dc} < J_{ON}$. The inset shows an example of the FMR-spectrum.

We also calculate $V_{dc}$ in the D state in the presence of the weak microwave current density. Fig. 2 shows an example of $V_{dc}$ as a function of frequency of the ac current calculated from Eq. 2 for direct current density $J_{dc}= 0.43\times10^8$ A/cm$^2$ and $J_M= 0.1\times10^8$ A/cm$^2$. In our systematic study as a function of field and direct current density, the maximum generated direct voltage in the D state is 4.5 $\mu V$ (at $J_{dc}= 0.35\times10^8$ A/cm$^2$) (one order of magnitude larger than the maximum ST-FMR voltage in the AP state).

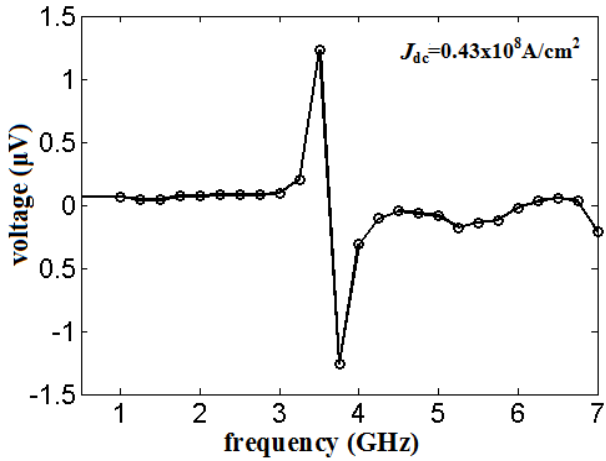

FIG. 2: Direct voltage as a function of frequency of the ac current in the D state calculated for $J_{dc}=0.43\times10^8$ A/cm$^2$ and $J_M= 0.1\times10^8$ A/cm$^2$.

Finally, we study the effects of thermal fluctuations on the STT-driven magnetization dynamics for several temperatures $T$ in the $0K \leq T \leq 300K$ range, direct current density in the $J_{OFF} < J_{dc} < J_{ON}$ range and zero alternating current. Our simulations show that thermal noise induces random jumps between the AP and the D state for temperatures $T \geq 50K$. In other words, the nanomagnet becomes super-paramagnetic. We find that the crossover temperature between the thermally stable and the super-paramagnetic behaviours depends on the dc bias current density as in the experiments.[19] Fig. 3 shows examples of time traces of the normalized average x-component of the magnetization for (a) $J_{dc}=0.38\times10^8$A/cm$^2$ at 150 K and (b) $J_{dc}=0.46\times10^8$A/cm$^2$ at 100 K.

In the superparamagnetic regime, for each value of the direct current density in the $J_{OFF} < J_{dc} < J_{ON}$ interval, random thermal torques induce asynchronous jumps between D and AP states governed by Kramer transition rates (diffusion over potential barriers) between a fixed point of dynamics (AP) and a limit cycle (D). Our data show that the dwell times in the D-state $\tau_D$ and in the AP state $\tau_{AP}$ strongly depend on $J_{dc}$. In particular, for $J_{dc}$ near $J_{ON}$, the free layer spends most of its time in the D state, while for $J_{dc}$ near $J_{OFF}$ the free layer spends most of its time in the AP state. Because one of the attractors is a limit cycle that introduces another time scale into the problem (the limit cycle period),[40] a quantitative analysis of thermally-induced transitions out of the D state is more complicated than that out of a fixed point.[41,42,43]

As we will discuss later in the text, for the bias current density and the temperature where the condition for the average dwell time $<\tau_D> \approx <\tau_{AP}>$ is achieved, ASR can be excited.

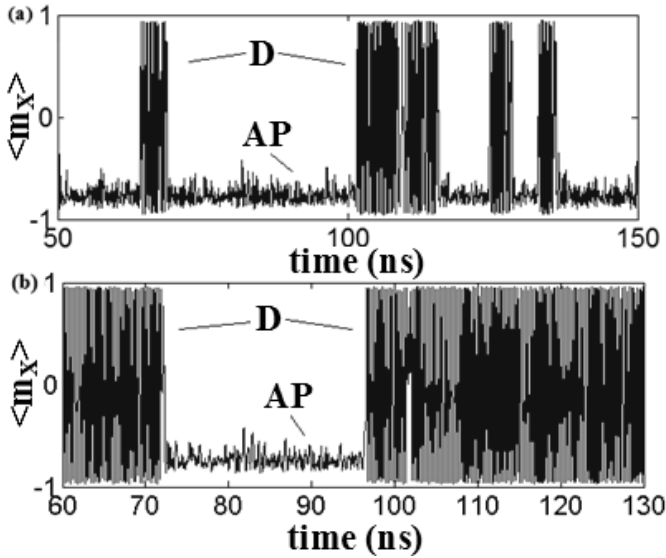

FIG. 3: Time traces of the normalized average x-component of the magnetization for (a) $J_{dc}$=0.38×10$^8$A/cm$^2$ at 150 K and (b) $J_{dc}$=0.46×10$^8$A/cm$^2$ at 100 K.

## IV. SIMULATIONS OF STOCHASTIC RESONANCE

In this section, we study the behaviour of the system related to transitions between a dynamic state D and static states in the presence of temperature and ac STT.

*a) Adiabatic stochastic resonance*

First, we briefly discuss our results on the magnetization dynamics in response to a low-frequency ac STT drive. We find that the low-frequency stochastic resonance or ASR is observed when the so called "time scale matching condition" is achieved.[21] Qualitatively, without ac STT drive the procedure to identify the direct current and temperature is based on the analysis of the average dwell times in the AP and D states. When the average dwell times in the AP and D states are nearly equal, the ac STT can excite ASR. For example at $T = 100$ K the $<\tau_D> \approx <\tau_{AP}>$ occurs at $J_{ASR}=0.5(J_{ON}+ J_{OFF}) = 0.43\times10^8$ A/cm$^2$. For these parameters, Fig. 4 displays the time trace at the ASR condition ($x$, $y$ and $z$ component of the average component of the magnetization) in presence of ac STT drive with $J_M= 0.1\times10^8$ A/cm$^2$ and $f_{AC}=50$MHz. This figure clearly demonstrates that the dynamics becomes quasi-periodic with the frequency of the ac drive (ASR dynamics). Quantitative analytical computations of the "time scale matching condition" in this system depends on the estimate of the effective potential energy curvature of the D-state. To do that for our dissipative system far from equilibrium, it is necessary to introduce a generalized effective potential that depends on magnetic anisotropy energy, STT and damping of the system[36,44]. Such a procedure goes beyond the scope of this paper and it will be implemented and discussed elsewhere.

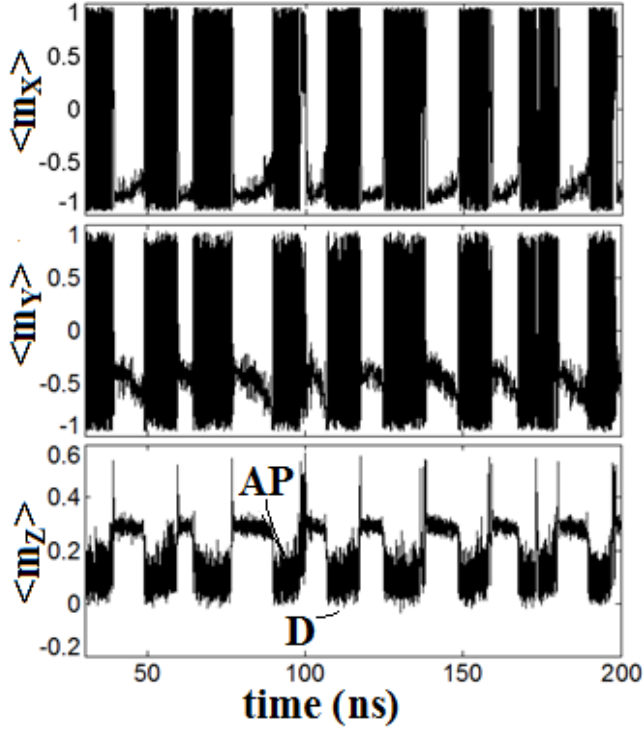

FIG. 4: Example of time traces (x, y and z average component of the magnetization) corresponding to ASR ($J_{dc}=0.43\times10^8$A/cm$^2$, $T=100$ K, $J_M=0.1\times10^8$ A/cm$^2$ and $f_{AC}=50$MHz).

*b) Non-adiabatic stochastic resonance*

It has been analytically predicted in under-damped bi-stable systems that in addition to the low-frequency ASR, a different non-linear resonant phenomenon, the NASR can be observed.[45,46,47] This kind of dynamical behaviour has been recently observed in nanoscale exchange biased spin valves.[19] In the rest of this paper, we show that micromagnetic simulations are able to reproduce the main experimentally observed features of NASR. Our micromagnetic simulations reveal the physical origin of NASR in the spin valve system.

Fig. 5(a) shows the resonant response of the dc voltage (calculated micro-magnetically using Eq. 2) as a function of the ac STT frequency for two current densities and temperatures $J_{dc}=0.35\times10^8$A/cm$^2$ ($T=100$K) and $J_{dc}=0.39\times10^8$A/cm$^2$ ($T=50$K). The simulation time is 500μs, which is long enough to calculate the dc voltage to high precision (see Fig. 6 where the time evolution of the dc voltage for $J_{dc}$=0.39×10$^8$A/cm$^2$, $T$=100K, $J_M$ =0.1×10$^8$A/cm$^2$ and $f_{AC}$=2.75GHz is shown). In agreement with the experiments, the simulations reveal existence of a high-frequency resonance (NASR) for which the dc voltage is much larger than the voltage calculated without thermal fluctuations (Fig. 1 and Fig. 2). More precisely, the induced dc voltage is two order of magnitude larger (the maximum calculated voltage is ~ 30 μV) than the dc voltage calculated in the AP state at $T$=0 and one order of magnitude larger than the voltage calculated in the D state at $T$=0 (compare Fig.5(a) to Fig.(2)).

To deeper understand the origin of this large dc voltage, we consider the time domain traces for different $f_{AC}$. Fig. 5(b) displays examples of the time traces of the average x-component of the magnetization for $J_{dc}$ = 0.35×10$^8$ A/cm$^2$ (T = 100 K) at two drive frequencies $f_{AC}$ = 2.75 GHz (top) and $f_{AC}$ = 6 GHz (bottom). Our results show that far from the resonance frequency, the magnetization spends most of its time in the AP state. However, near the resonance frequency, the magnetization exhibits hopping among the AP state, the D state and the parallel state of the spin valve (P state) with the dwell times significantly exceeding $1/f_{AC}$. Fig. 7(a) shows the trajectory of the average magnetization vector at NASR. For comparison, zero-temperature trajectories in the D-state and the AP-state are also shown for $f_{AC}$ = 2.75 GHz in Fig. 7(b). In contrast to the ASR where only the single-path D→AP and AP→D transitions take place, at NASR we observe transitions among all three states: AP, P and D, but we do not observe a direct transition D→AP, only an indirect D➔P➔AP transition takes place. The transition among those states are mainly due to magnetization rotation with quasi-uniform configuration of the magnetization.[48] Our simulations suggest that a regular finite state Markov chain process gives an appropriate description of the dynamics at NASR. A Markov process is a stochastic process which has the property that the probability of a transition from a given state A to a future state B is dependent only on the present state and not on the manner in which the current state A was reached. A finite state Markov chain describes a sequence of Markov-process transitions among a finite number of discrete states (P, AP and D states for our system).

Equation 3 shows the transition matrix **M** (stochastic matrix) of the Markov chain characterized for the time domain data of Fig.5b (top), each element of the matrix represents the probability that the magnetization jumps from one state (left of the matrix) to another (top of the matrix) in one step, for example $a_{12}$ is the probability of the magnetization transition from the P-state to the AP-state, for example $a_{12} = \frac{n_{T,P \to AP}}{n_{T,P \to AP} + n_{T,P \to D}}$ being $n_{T,P \to AP}$ and $n_{T,P \to D}$ the number of transitions from the P-state to the AP-state and the D-state respectively.

$$\begin{matrix} & [P \quad AP \quad D] \\ \begin{bmatrix} P \\ AP \\ D \end{bmatrix} & \begin{pmatrix} 0 & a_{12} & a_{13} \\ a_{21} & 0 & a_{23} \\ a_{31} & 0 & 0 \end{pmatrix} \end{matrix} = \mathbf{M} = \begin{pmatrix} 0 & 0.75 & 0.25 \\ 0.35 & 0 & 0.65 \\ 1 & 0 & 0 \end{pmatrix}_{f=2.75 GHz} \quad (3)$$

Each element of the transition matrix is calculated as a statistical average over the entire ensemble of transitions observed in the simulation.

Table I summarizes the time scales for each state (for example, $<\tau_{AP \to P}>$ is the average time the magnetization spends in the AP state before the transition to the P state is achieved).[42] The system shows a different behaviour far from the NASR frequency, in particular the magnetization spends most of the time in a single state, and thus the regular Markov chain seen at NASR becomes an absorbing Markov chain.[49]

| Transition | $<\tau>$ (ns) |
|---|---|
| P→AP | 0.53 |
| P→D | 1.86 |
| D→P | 2.42 |
| D→AP | NO |

| AP→D | 6.87 |
| AP→P | 4.53 |

TABLE I. Characteristic time scale for each transition.

Quantitatively, at the generated dc voltage has two components:

$$V_{dc} = I_{ac}\delta R_{ac} + I_{dc}\delta R_{dc} \quad (4)$$

The first term originates from the resistance oscillations at the drive frequency, $\delta R_{ac}$, in the AP, P and D states. The ac STT drive excites oscillations of magnetization which give rise to oscillations of the resistance at the drive frequency with the amplitudes $\delta R_{ac-AP}$, $\delta R_{ac-P}$, and $\delta R_{ac-D}$ respectively. The global $\delta R_{ac}$ is given by:

$$\delta R_{ac} = a_{ac-AP}\delta R_{ac-AP} + a_{ac-P}\delta R_{ac-P} + a_{ac-D}\delta R_{ac-D} \quad (5)$$

where $a_{ac-P}$, $a_{ac-AP}$, and $a_{ac-D}$ are coefficients that depend on the phase shift between the resistance and current oscillations in the P, AP and D states [50] as well as on the average dwell times in these states. The term proportional to $I_{dc}$ is related to the time-average resistance variation, $\delta R_{dc}$, in the presence and the absence of the ac drive. This term originates from ac-driven hopping among P, AP and D states that have different time-average resistance values. The first term in Eq. (4) is dominant when the magnetization is mainly in a single state $I_{dc}\delta R_{dc} \approx 0$, while the second term is dominant if the ac drive excites hopping among different states and for a different frequency the magnetization is manly in a single state. To compare our data directly to the experiments of Refs. [19, 20], we systematically study the maximum dc voltage ($V_{DC-max}$) at the NASR frequency as a function of the direct current density and temperature. Fig. 5(c) shows a comparison between the $V_{DC-max}$ as function of the direct current density for $T$=100K and $T$=0K. We observe a significant NASR-generated voltage for direct current densities between $J_{ON}$ and $J_{ASR}$ where the second term in Eq. 4 is dominant. As the current density continues to increase, the $V_{DC-max}$ approaches the value calculated at $T$=0K because the magnetic configuration is mainly in the D state and the first (smaller) term of Eq. 4 is dominant (the same as at $T$=0K). In this context, the D-state acts as absorbing state in the absorbing Markov chain. Fig. 5(d) displays the temperature dependence (step of 50K) of the $V_{DC-max}$ for three different direct current densities $J_{dc}$ = 0.35, 0.39, and 0.43×10$^8$A/cm$^2$ (an offset of 10 μV is applied for each curve). This non-monotonic dependence of the resonance amplitude on temperature is typical for stochastic resonance phenomena[21] and is in a good agreement with the experiment.[19]

Our calculations reproduce all major features of the NASR effect recently observed in nanoscale spin valves[19]. First, for a fixed temperature, $V_{DC-max}$ shows a non-monotonic behaviour as a function of the direct current density (see Fig. 3(b) of Ref. [19]). Second, the curves of $V_{DC-max}$ show non-monotonic temperature dependence typical for stochastic resonance (see Fig. 3(d) of Ref.[19]). Third, the largest value of $V_{DC-max}$ shifts to lower temperatures with increasing direct current density (see Fig. 3(f) of Ref. [20]).

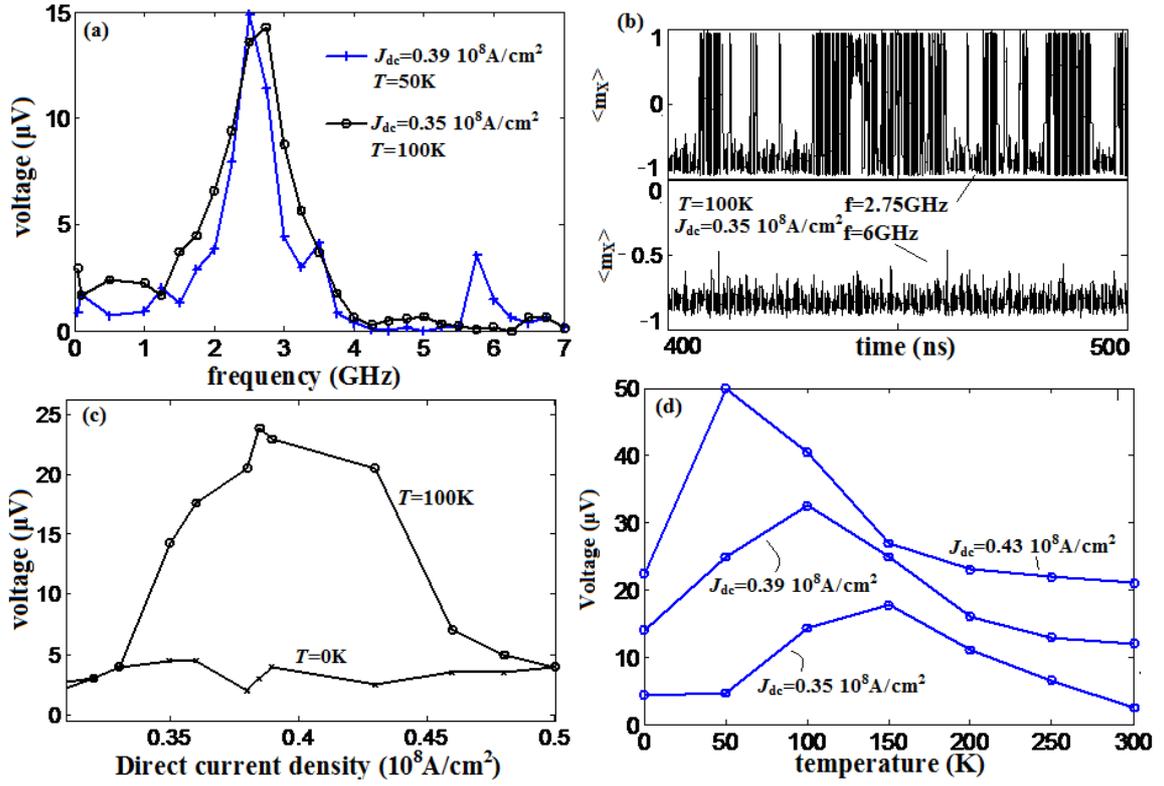

FIG. 5: (Color online) (a) Dc voltage as function of frequency of the applied alternating current calculated for two direct current densities and temperatures, $J_{dc}=0.35\times10^8 A/cm^2$ ($T=100K$) and $J_{dc}=0.39\times10^8 A/cm^2$ ($T=50K$); (b) time traces of the average x-component of the magnetization for $J_{dc}=0.35\times10^8 A/cm^2$ ($T=100K$) at two different ac current frequencies, $f_{AC}=2.75GHz$ (top) and $f_{AC}=6GHz$ (bottom); (c) $V_{DC-max}$ as function of the direct current density calculated at $T=100K$ and $T=0K$; (d) Temperature dependence of $V_{DC-max}$ for three direct current densities, $J_{dc}=0.35$, $0.39$, and $0.43\times10^8 A/cm^2$ (the curves are vertically offset by 10 μV for clarity).

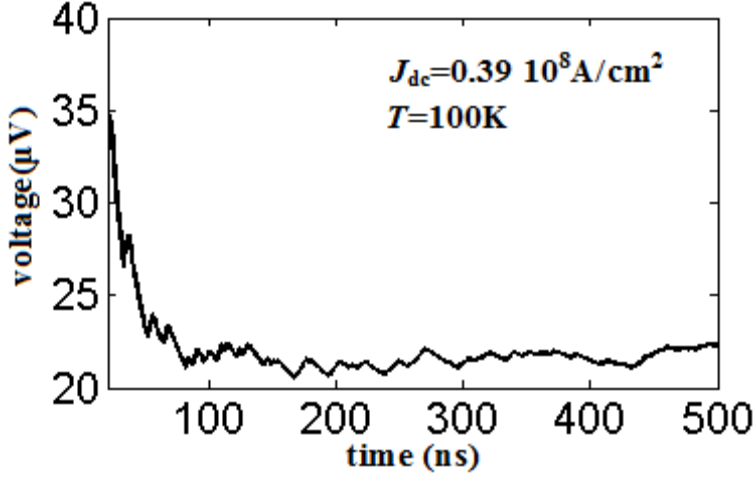

FIG. 6: Time evolution of dc voltage computed for $J_{dc}=0.39\times10^8$A/cm$^2$, $T$=100K, $J_M=0.1\times10^8$A/cm$^2$ and $f_{AC}$=2.50 GHz.

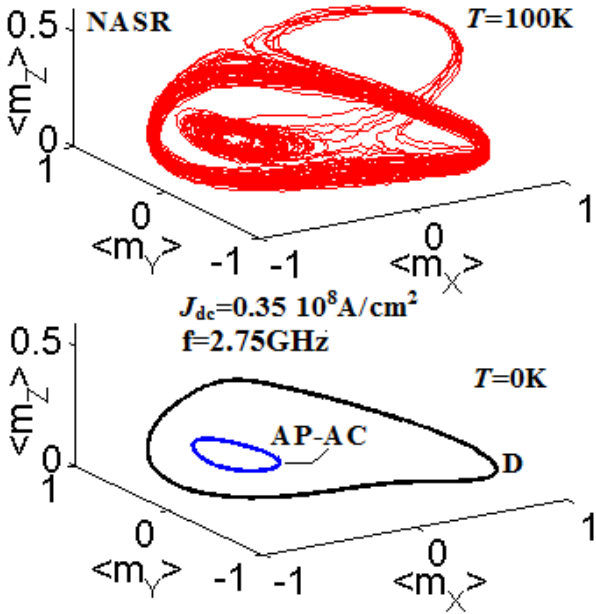

FIG. 7: (Color online)(a) Trajectory of the average magnetization vector at the non-adiabatic stochastic resonance (NASR) for $J_{dc}=0.35\times10^8$A/cm$^2$, $T$=100K, $J_M=0.1\times10^8$A/cm$^2$ and $f_{AC}$=2.75GHz. (b) Zero-temperature magnetization trajectories in the D and the AP (AP-AC) states for the drive frequency of $f_{AC}$=2.75GHz.

*c) Example of application of non-adiabatic stochastic resonance*

The NASR driven by STT can be used in applications such as microwave signal detection. Fig. 8 shows a simplified schematic block diagram of a receiver for the frequency-shift keying[51] (FSK) modulation based on the STT-NASR. The microwave signal to be demodulated, $s_M(t)$, is applied to the two non-adiabatic stochastic resonator receivers which are designed to have different resonance frequency. Depending on the frequency of the $s_M(t)$, the detector will identify by means of the output voltage which of the resonators is active and will consequently decode the original bit encoded in $s_D(t)$. The system can be also generalized for a M-ary FSK[51] by introducing M non-adiabatic stochastic resonators. From the practical point of view, the stochastic resonator receiver can be also realized using a magnetic tunnel junction (MTJ)[52] instead of a giant magnetoresistance spin valve considered here in order to have much larger output voltage. In particular, recent experiments report NASR dc voltage values as high as 30 mV for ac current densities as low as ~$10^5$ A/cm$^2$ in MTJ NASR resonators.[53]

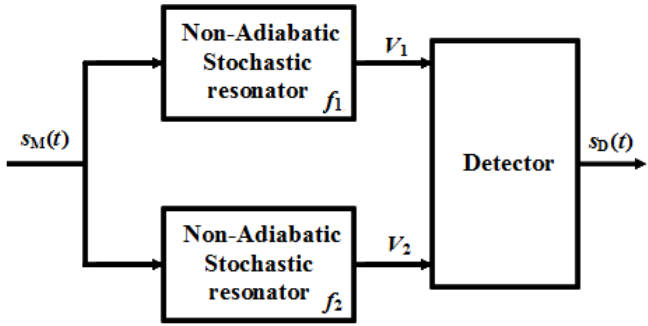

FIG. 8: A simplified schematic block diagram of a receiver for frequency-shift keying (FSK) modulation based on non-adiabatic stochastic resonance phenomenon.

## IV. SUMMARY AND CONCLUSIONS

We employed micromagnetic simulations to study the phenomenon of stochastic resonance excited by spin transfer torque and thermal fluctuations in nanoscale exchange biased spin-valves. In agreement with experiment, our simulations reveal the existence of both the low-frequency (adiabatic) and a high-frequency (non-adiabatic) stochastic resonance effects. The ASR, existing in a narrow range of direct current densities around $J_{ASR}$, is observed in the low frequency regime ($f < 500$MHz) and it is characterized by quasi-periodic transitions between a static (AP) state and a dynamic (D) state of magnetization auto-oscillations. The NASR is observed at a high frequency of the ac STT drive. It is achieved in a wider direct current density range and it involves transitions among two static (P and AP) and one dynamic (D) states with transition probabilities describable by a regular Markov chain. The direct voltage generated by the spin valve at NASR is enhanced by one order of magnitude compared to the direct voltage generated at zero temperature. The results of our simulations are in excellent agreement with recent measurements of stochastic resonance driven by spin torque in nanoscale spin valves. We also propose a simple design of a sensitive microwave signal detector based on NASR driven by spin transfer torque.

## ACKNOWLEDGMENTS

This work was supported by Spanish Projects under Contracts No. SA025A08 and MAT2008-04706/NAN, by NSF Grants No. DMR-0748810 and No. ECCS-0701458, by DARPA Grant No. HR0011-09-C-0114 and by the Nanoelectronics Research Initiative through the Western Institute of Nanoelectronics.